\documentclass{PoS}
\usepackage{pstricks,pst-eps,pst-plot,pst-node,pst-text,pst-3d,pst-coil,pst-grad}
\usepackage{graphicx}
\usepackage{amsmath}
\usepackage{amssymb}
\usepackage{longtable}

\PoS{PoS(LAT2005)092}
\newcommand{\no}{\nonumber}

\newcommand{\be}{\begin{eqnarray}}
\newcommand{\ee}{\end{eqnarray}}
\newcommand{\bq}{\begin{equation}}
\newcommand{\eq}{\end{equation}}
\newcommand{\ba}{\begin{array}}
\newcommand{\ea}{\end{array}}
\newcommand{\la}{\langle}
\newcommand{\ra}{\rangle}
\newcommand{\dsla}{D \hspace{-1.7ex}/~}

\newcommand{\partsla}{\partial \hspace{-7pt}/~}

\newcommand{\vsa}{\rlap{\hbox{$\longrightarrow$}}
                   \raise 7pt \hbox{\scriptsize VSA~~}}
\newcommand{\ope}{\rlap{\hbox{$\Longrightarrow$}}
                   \raise 2ex \hbox{\hspace{+0.1ex}\scriptsize OPE~~}}

\newcommand{\nabr}{\rlap{\hbox{$\nabla$}}
                   \raise 8 pt \hbox{$\hspace{-0.05cm}\leftarrow$}}
\newcommand{\nabd}{\rlap{\hbox{\boldmath $D$}}
                   \raise 15 pt \hbox{$\hspace{-0.05cm}\leftarrow$}}
\newcommand{\rvD}{\rlap{\hbox{$\vec{D}$}}
                   \raise 8 pt \hbox{$\hspace{-0.05cm}\leftarrow$}}
\newcommand{\dslar}{\rlap{\hbox{$\dsla$}}
                   \raise 8 pt \hbox{$\hspace{-0.05cm}\rightarrow$}}
\newcommand{\dslal}{\rlap{\hbox{$\dsla$}}
                   \raise 8 pt \hbox{$\hspace{-0.05cm}\leftarrow$}}
\newcommand{\dl}{\rlap{\hbox{$D_\mu$}}
                   \raise 8 pt \hbox{$\hspace{-0.05cm}\leftarrow$}}
\newcommand{\partr}{\rlap{\hbox{$\partial_\mu$}}
                   \raise 8 pt \hbox{$\hspace{-0.05cm}\reftarrow$}}
\newcommand{\partl}{\rlap{\hbox{$\partial_\mu$}}
                   \raise 8 pt \hbox{$\hspace{-0.05cm}\leftarrow$}}
\newcommand{\partslar}{\rlap{\hbox{$\partsla$}}
                   \raise 8 pt \hbox{$\hspace{-0.05cm}\rightarrow$}}
\newcommand{\partslal}{\rlap{\hbox{$\partsla$}}
                   \raise 8 pt \hbox{$\hspace{-0.05cm}\leftarrow$}}

\newcommand{\simg}{\rlap{\raise -4pt \hbox{$\sim$}}
                   \raise 3pt \hbox{$>$}}
\newcommand{\siml}{\rlap{\raise -4pt \hbox{$\sim$}}
                   \raise 3pt \hbox{$<$}}

\newcommand{\smf}{\rlap{\hbox{$\ \longrightarrow$}}
                   \raise 7pt \hbox{\scriptsize $am_f\ll 1$}}
\newcommand{\ninf}{\rlap{\hbox{$\ \longrightarrow$}}
                   \raise 7pt \hbox{\scriptsize $N_5\rightarrow\infty$~~}}

\def\mKn{{m_{K^0}}}
\def\mKc{{m_{K^+}}}

\def\alpem{\alpha_{\rm em}}

\title{
Electromagnetic properties of hadrons with two flavors of dynamical
domain wall fermions\footnote{
We thank RIKEN, Brookhaven National Laboratory and the U.S. Department
of Energy for providing the facilities and hospitality where this work
was done.}}

\ShortTitle{
Electromagnetic properties of hadrons with two flavors of dynamical
domain
$\cdots$
}

\author{\speaker{Norikazu~Yamada}
        \\
        High Energy Accelerator Research Organization(KEK), Tsukuba,
        Ibaraki 305-0801, Japan \\
        The Graduate University for Advanced Studies, Tsukuba,
        Ibaraki 305-0801, Japan \\
        E-mail: \email{norikazu.yamada@kek.jp}}
\author{Thomas~Blum\\
        Physics Department, University of Connecticut, Storrs, CT
        06269-3046, USA\\
        RIKEN BNL Research Center, Brookhaven National Laboratory,
        Upton, New York 11973, USA\\
        E-mail: \email{tblum@phys.uconn.edu}}
\author{Masashi~Hayakawa\\
        Theoretical Physics Group, RIKEN, Wako 2-1, Saitama 351-0198,
        Japan\\
        E-mail: \email{haya@postman.riken.go.jp}}
\author{Taku~Izubuchi\\
        Institute for Theoretical Physics, Kanazawa University, Kanazawa
        920-1192, Japan\\
        RIKEN BNL Research Center, Brookhaven National Laboratory,
        Upton, New York 11973, USA\\
        E-mail: \email{izubuchi@quark.phy.bnl.gov}}
\author{for the RBC Collaboration}
\date{\today}
\abstract{
 We present the determination of the light quark masses using
 electromagnetic splittings of pions and kaons.
 The meson masses are calculated on $SU(3)\times U(1)$ gauge
 configurations, where $SU(3)$ gauge fields include sea quark effects of
 two degenerate flavors and the $U(1)$ gauge fields are incorporated in
 non-compact form. Possible sources of systematic uncertainty are discussed.
}

\FullConference{XXIIIrd International Symposium on Lattice Field Theory\\
		 25-30 July 2005\\
		 Trinity College, Dublin, Ireland}

\begin{document}

\section{Introduction}
\label{sec:intro} 

Electromagnetic (EM) properties of hadrons offer a rich source of
interesting and important phenomena.
The patterns of the mass splittings between charged and neutral mesons
or the mass splittings among the octet or decuplet baryons are sensitive
to the isospin breaking of up and down quark masses and the presence of
EM interactions. It is also known that the width difference of the $\rho^+$ and $\rho^0$
mesons and
the hadronic light-by-light scattering amplitude, which may not be
computed from measured EM properties of hadrons, play an important role
in the standard model (SM) prediction of the anomalous magnetic moment
of muon.

Recent developments in lattice QCD in both hardware and
software have advanced the field closer to the goal of QCD 
calculations without approximation; large-scale, high precision unquenched simulations
are becoming available~\cite{Izubuchi:lattice2005 review}.
What is important here is that the statistical error of pseudoscalar meson
masses are well under control, with size typically a few
percent or less.
Remembering that charged-neutral meson splittings are of
$O(\alpha_{\rm em})\sim O(1\%)$, where $\alpha_{\rm em}$ is the fine
structure constant, it is expected that once EM interactions are successfully included, 
it will be possible to determine the
up and down quark masses, which are the poorly known parameters of the
SM, from first principles by using such splittings as inputs, and
thus we can check the simplest solution to the strong CP problem,
$m_u$=0.

In this talk, we focus on the determination of the light quark masses.
The strategy is basically following
Refs.~~\cite{Duncan:1996xy}, where the EM fields are introduced in a
non-compact form, and combined with QCD link variables to realize QCD +
QED theory on the lattice.
While this pioneering work made use of the Wilson quark action in the
quenched approximation to QCD, here we employ domain wall
fermions~\cite{Kaplan:1992bt,Shamir:1993zy} and QCD
configurations with two flavors of dynamical domain wall fermions,
generated by the RBC Collaboration~\cite{Aoki:2004ht}.
Our activity toward the lattice calculation of the anomalous magnetic
moment of muon is reported in Ref.~\cite{Hayakawa:2005eq}.

\section{QCD+QED system}
\label{sec:nc-qed}

We employ unquenched QCD gauge configurations $U_{\rm qcd,\it\mu}(x)$,
generated with parameters $V=16^3\times32$, $L_s=12$, $M_5=1.8$
and three different sea quark masses, $m_{\rm sea}=0.02$, $0.03$,
$0.04$~\cite{Aoki:2004ht}.
The lattice spacing is determined to be $1.691(53)$ GeV from
$m_\rho$=770 MeV, and hence the physical volume is roughly
(1.9 fm)$^3$.
Calculations are performed on about 200 configurations from the 5,000 trajectories 
available at each value of the sea quark mass.

A non-compact form is adopted for the action of the $U(1)$ gauge fields,
$A_{\rm em,\it\mu}(x)$~\cite{Duncan:1996xy}.
To generate $A_{\rm em,\it\mu}(x)$, we first rewrite the action in
momentum space, and impose the Coulomb gauge condition.
After diagonalizing the kernel, the gauge fields are chosen
randomly according to a Gaussian distribution and with $e$=1.
The configuration $A_{\rm em,\it\mu}(x)$
is then obtained by Fourier transformation.
It is worth noting that with the use of the non-compact action and the
generation procedure described above there is no auto-correlation among
the configurations, even for arbitrarily small coupling.
Since quenched QED is a free theory, the fine structure constant
$\alpha_{\rm em}$ receives no renormalization.
Using $A_{\rm em,\it\mu}(x)$ and the quark electric charge $Q_q$,
we obtain a $U(1)$ link variable,
$({U_{\rm em}}_\mu(x))^{Q_q} = e^{i Q_q A_{\rm em,\it\mu}(x)}$.
Configurations for the (QCD + QED) theory are then constructed by
$U_{\rm qcd,\it\mu}(x)\times (U_{\rm em,\it\mu}(x))^{Q_q}$, which are
used in the inversion of the Dirac operator for valence quarks.
Though $e$=1 is adopted in the configuration generation, any value of
the quark's electric charge can be realized by tuning $Q_q$.

In this system, the flavor non-singlet axial Ward-Takahashi (WT)
identity for two flavors of domain wall fermions with mass $m_{q1}$ and
$m_{q2}$ and charge $Q_{q1}$ and $Q_{q2}$ is given by
\begin{equation}
    \partial_{\mu}^* {\cal A}^a_\mu(x)
=  (m_{q1}+m_{q2})\ P^a(x)
   + (m_{q1}-m_{q2})\ \bar{q}(x) \gamma_{5}
     \{\frac{\tau^a}{2},\frac{\tau^3}{2}\}\,q(x)
   + 2\,J_{5q}^a(x)
   + \sum_s\epsilon(s){\cal X}_s^a(x),
  \label{eq:w-t iden}
\end{equation}
where ${\cal A}_\mu^a(x)$ takes the same form as the conserved
axial-vector current in the pure QCD case, but with link variables,
$U_{\rm qcd,\it\mu}(x)\times$
diag.$[(U_{\rm em,\it\mu}(x))^{Q_{q1}},
       (U_{\rm em,\it\mu}(x))^{Q_{q2}}]$,
$P^a(x)$ a pseudoscalar density, $J_{5q}^a(x)$ the same as in pure QCD,
and
\begin{eqnarray}
 \hspace*{-5ex}
&&  {\cal X}_s^a(x)
 = -\frac{1}{2}\sum_\mu\Bigg[\ \
     \bar\Psi_s(x)
     (1-\gamma_\mu)U_{\rm qcd,\it x,\mu}\,
           \left( (U_{\rm em,\it\mu}(x))^{Q_{q1}}
                - (U_{\rm em,\it\mu}(x))^{Q_{q2}}
           \right)\,[\frac{\tau^a}{2},\frac{\tau^3}{2}]
     \Psi_{s}(x+\hat\mu)\no\\
\hspace*{-5ex}&&
    + \bar\Psi_s(x)
       (1+\gamma_\mu)U^\dag_{\rm qcd,\it\mu}( x-\hat\mu)\,
        \left( (U^{\dagger}_{\rm em,\it\mu}(x-\hat\mu))^{Q_{q1}}
             - (U^{\dagger}_{\rm em,\it\mu}(x-\hat\mu))^{Q_{q2}}
       \right)\,[\frac{\tau^a}{2},\frac{\tau^3}{2}]
      \Psi_s(x-\hat\mu)
                 \Bigg],
\label{eq:X}
\end{eqnarray}
and $\epsilon(s)=1$ for $1\le s \le L_s/2$ and $-$1 for $L_s/2 <  s \le L_s$.
Notice that ${\cal X}_s^a(x)$ vanishes when $a=3$ or $Q_{q1}=Q_{q2}$.

From the analogy to the fact that in pure QCD with the domain-wall
formalism the presence of the singlet axial anomaly becomes clear by
considering the disconnected diagram involving the singlet, mid-point
current, it is inferred that the similar understanding is possible for
the $U(1)$ anomaly, but this time the disconnected diagram involving
the flavor non-singlet, mid-point current $J^a_{5q}(x)$ plays an
important role.
Due to the presence of the $U(1)$ anomaly, none of the components of
${\cal A}_\mu^a(x)$ is conserved for $Q_{q1}\ne Q_{q2}$ even in the quark
massless limit, and hence there is no Nambu-Goldstone (NG) boson in this
system~\cite{Suganuma}.
If you consider the lattice measurement of $m_{\pi^0}$ through the
standard method using two-point correlation functions, the possible
origin of the non-vanishing $m_{\pi^0}$ is the $\pi^0$-singlet mixing,
and it could affect $m_{\pi^0}$ by $O(\alpha_{\rm em}^2)$ or higher.
At present, we do not include the $\pi^0$-singlet mixing.
Thus, $\pi^0$ must become massless in the quark massless limit.
Notice that even after including the mixing effects, it is justified to
regard $\pi^0$ as an NG boson up to $O(\alpha_{\rm em}^2)$.
In addition to the above approximation, we are neglecting the
disconnected diagram in $\pi^0$-$\pi^0$ correlation function (for now).
It should be noted that the latter approximation does not affect the
fact that ``$\pi^0$'' becomes massless in the quark massless limit, and
more importantly, the point where $m_{\pi^0}$ vanishes.

\section{Numerical results}
\label{sec:general remarks}

At present, we only have meson data in which the two valence quarks are
degenerate with the sea quark mass.
The electric charge of the valence quarks is set to $Q_q=+2e/3$ for the up
quark and $-e/3$ for the down quark.
In this work, we take three values of $e$, which correspond to
$\alpha_{\rm em} = \alpha_{\rm em,\,phys},
\frac{(0.6)^2}{4\pi},\,\frac{(1.0)^2}{4\pi}$, to examine the
$\alpha_{\rm em}$ dependence of the meson mass splittings.
The mass of the pions is extracted from two-point correlation functions
as usual.
We employ two different operators, $P^a(x)$ and $A^a_0(x)$, as the
interpolating operator.
Since the EM interactions violate isospin symmetry, $\pi^0$
mixes with the lightest isosinglet state.
However, this mixing turns out to start only at
$O(\alpha_{\rm em}^2)$, and hence we neglect this effect as the present
statistics cannot resolve such a small effect.
Although the calculation of $m_{\pi^0}$ requires a disconnected diagram,
we do not include it, as mentioned earlier. We will come back to this point later.

At the leading order of chiral perturbation theory (ChPT) including the
EM interactions, $m_{\pi^+}^2$ and $m_{\pi^0}^2$ are given by
\begin{eqnarray}
 m_{\pi^+}^2=&\alpha_{\rm em}\,\Delta_+^{(0)}+&
             2\,(B_0 +\alpha_{\rm em}\,\Delta_+^{(m)})\,
              (m_{f,+}+ m_{\rm res}),
              \label{eq:chpt pi+}\\
 m_{\pi^0}^2=&& 2\,( B_0 +\alpha_{\rm em}\,\Delta_0^{(m)})\,
              (m_{f,+}+ m_{\rm res}),
              \label{eq:chpt pi0}
\end{eqnarray}
to $O(\alpha_{\rm em}^2)$, where $m_{f,\pm}=(m_{f,d}\pm m_{f,u})/2$.
In the following, these functional forms are used to determine the low
energy constants (LEC's), $B_0$, $\Delta_+^{(0)}$, $\Delta_+^{(m)}$ and
$\Delta_0^{(m)}$.
$m_{\rm res}$ is estimated by finding $m_f$ at which $m_{\pi^0}^2$
vanishes, since $\pi^0$ can be considered as an NG boson.

\begin{figure}
 \centering
 \includegraphics*[width=0.49 \textwidth,clip=true]
 {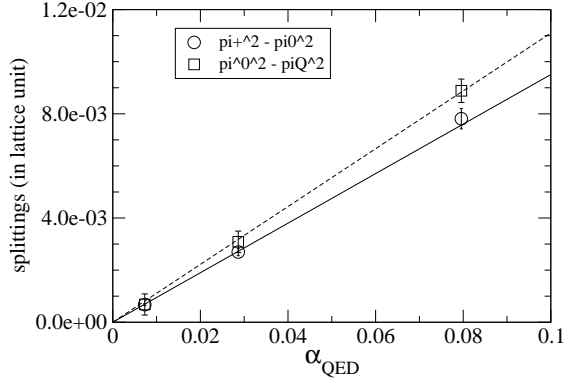}
 \caption{The $\alpha_{\rm em}$ dependence of the splittings.
 The results from $\la A_0A_0 \ra$ are shown.}
 \label{fig:e-dep of splittings}
\end{figure}

Before discussing the determination of the LEC's, let us mention
$O(\alpha_{\rm em})^2$ corrections.
Figure~\ref{fig:e-dep of splittings} shows the $\alpha_{\rm em}$
dependence of the $\pi^+$-$\pi^0$ and $\pi^0$-$\pi^Q$ mass splittings, where
$\pi^Q$ denotes the pion in pure QCD.
It is seen that each splitting is well described by a linear function
of $\alpha_{\rm em}$ in the range $\frac{1}{137}< \alpha_{\rm em} < 0.08$.
This means that $O(\alpha_{\rm em}^2)$ contributions are not significant
in this range of $\alpha_{\rm em}$.
This observation allows us to treat the EM interactions perturbatively.

\begin{figure}
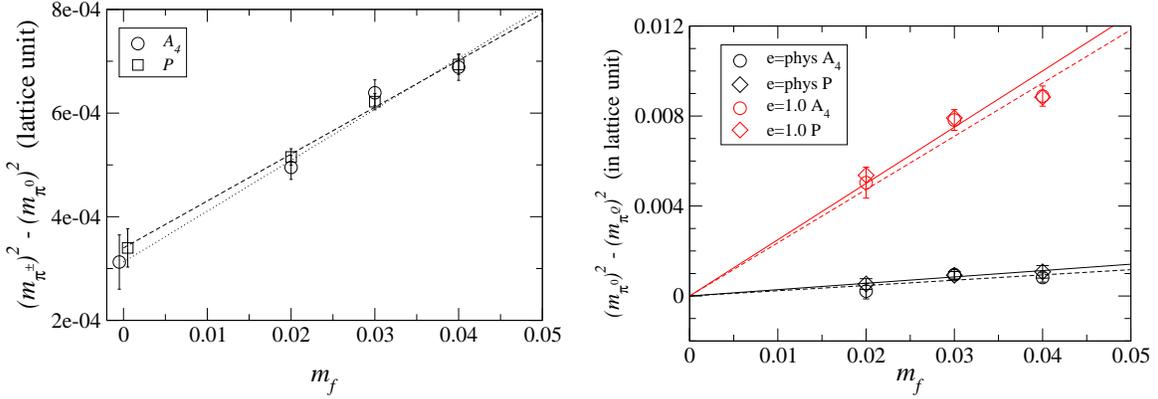

 \centering
 \begin{tabular}{cc}
   \includegraphics*[width=0.49 \textwidth,clip=true]
   {Figs/squ_pi+-pi0_9-23_ephys.eps} &
   \includegraphics*[width=0.49 \textwidth,clip=true]
   {Figs/squ_pi0-piQ_9-23.eps}\\
 \end{tabular}
 \caption{The $m_f$ dependence of $\pi^+$-$\pi^0$ splitting (left) and
 $\pi^0$-$\pi^Q$ splitting (right).}
 \label{fig:splittings}
\end{figure}

Now we proceed to the determination of the LEC's. The entire analysis is done under a
jack-knife procedure.
$B_0$ is obtained by fitting the pure-QCD pion data, $m_{\pi^Q}^2$.
By fitting $m_{\pi^+}^2-m_{\pi^0}^2$ as a function of $m_f$, 
the combination $(\Delta_+^{(m)}+\Delta_0^{(m)})$ and $\Delta_+^{(0)}$ are
extracted.
Finally $\Delta_0^{(m)}$ is obtained from the slope of
$m_{\pi^0}^2-m_{\pi^Q}^2$.
In Fig.~\ref{fig:splittings}, the $m_f$ dependence of
$m_{\pi^+}^2-m_{\pi^0}^2$ and $m_{\pi^0}^2-m_{\pi^Q}^2$ and the
resulting fit curves are shown as examples.
From these fits we obtain
\begin{eqnarray}
       B_0 = 2.09(4),\ \
       \Delta^{(0)}_+ = 0.0429(71),\ \
       \Delta^{(m)}_+ = 2.28(31),\ \
       \Delta^{(m)}_0 = 1.60(28),\ \
       &&\mbox{from $A_0$},
       \label{eq:LEC A0}\\
       B_0 = 2.06(4),\ \
       \Delta^{(0)}_+ = 0.0466(51),\ \
       \Delta^{(m)}_+ = 2.56(29),\ \
       \Delta^{(m)}_0 = 1.94(27),
       &&\mbox{from $P$},
       \label{eq:LEC P}
\end{eqnarray}
in lattice units.
Since we have not measured masses of non-degenerate mesons, we simply apply
eqs.~(\ref{eq:chpt pi+}) and (\ref{eq:chpt pi0}) to estimate the kaon masses,
\begin{eqnarray}
 m_{K^+}^2=&\alpha_{\rm em}\,\Delta_+^{(0)}+&
            ( B_0 +\alpha_{\rm em}\,\Delta_+^{(m)})\,
              (m_{f,s}+ m_{\rm res,s}
              +m_{f,u}+ m_{\rm res,u}),\\
 m_{K^0}^2=&& ( B_0 +\alpha_{\rm em}\,\Delta_0^{(m)})\,
              (m_{f,s}+ m_{\rm res,s}
              +m_{f,d}+ m_{\rm res,d}).
\end{eqnarray}
Using the lowest order ChPT expressions,
\begin{eqnarray}
 \hspace*{-3ex}
 && m_{\pi^0}^2 =2\,\left(\, B_0
        + \alpha_{\rm em}\,\Delta_0^{(m)}\,\right)\,
    \left(m_{f,+}+ m_{\rm res,\it +}\right),\\
 \hspace*{-3ex}
 && (\mKc^2+\mKn^2)
        = 2\,{B_0}\,({m_{f,s}}+m_{f,+}
                     +m_{\rm res,\it s}+m_{\rm res,\it +})
        \no\\&&\hspace*{16ex}
        + \alpem\,
          \bigg( {\Delta^{(0)}_+}
              + \big( {\Delta^{(m)}_0} + {\Delta^{(m)}_+}\,\big)
                (m_{f,s}+m_{\rm res,\it s})
          \bigg),\\
 \hspace*{-3ex}
&&  ( m_{K^0}^2   - m_{K^+}^2   )
        -( m_{\pi^0}^2 - m_{\pi^+}^2 )
        = 2\,B_0\,({m_{f,-}}+m_{\rm res,\it -})
        + \alpem\,\big(
          \Delta^{(m)}_0
        - \Delta^{(m)}_+ \big)\,(m_{f,s}+m_{\rm res,\it s}),
\end{eqnarray}
the results of eqs.~(\ref{eq:LEC A0}) and (\ref{eq:LEC P}),
and the experimental values of the meson masses, 
we can determine $m_{f,\pm}$ and $m_{f,s}$ ($O(\alpha_{\rm em}m_{f,\pm})$ terms 
have been neglected).
Using $a^{-1}$=1.691 GeV, and the non-perturbative value of the quark mass renormalization factor
$1/Z_m=Z_S$=0.62~\cite{RBC NP Zm}
($m^{\overline{\rm MS}}(2 {\rm GeV})=Z_m(m_f+m_{\rm res})$), we obtain the
following preliminary results for the quark masses (statistical error only)
\begin{eqnarray}
&& m^{\overline{\rm MS}}_u(2 {\rm GeV},{\rm ref})
 = 2.84(23) \mbox{ MeV ($A_0$)},\ \ \
   2.89(8) \mbox{ MeV ($P$)},
   \label{eq:mu}\\
&& m^{\overline{\rm MS}}_d(2 {\rm GeV},{\rm ref})
 = 5.41(24) \mbox{ MeV ($A_0$)},\ \ \
   5.49(13) \mbox{ MeV ($P$)},
   \label{eq:md}\\
&& m^{\overline{\rm MS}}_s(2 {\rm GeV},{\rm ref})
 = 106.8(8) \mbox{ MeV ($A_0$)},\ \ \
   106.5(5) \mbox{ MeV ($P$)}.
   \label{eq:ms}
\end{eqnarray}
While the statistical errors appear to be under control,
we emphasize that our results are determined in the quenched 
approximation of QED, and we have used only $m_f =m_{val} = m_{\rm sea}$
data points in our fits (`ref' in the above results is stressing it).
It has been reported that in the pure QCD calculation partially-quenched
(PQ) valence quark mass effects at the next-to-leading order in ChPT
tend to increase the strange quark mass compared to the lowest-order
extrapolation used here~\cite{Aoki:2004ht}. In addition, the naive
definition of $m_{res}$ used in this analysis amounts to a constant
shift downward of a few percent on each of the quark masses compared to
the results in~\cite{Aoki:2004ht}.
Thus our results should be considered preliminary.
On the other hand, the effects of the EM interactions relative to pure QCD
(and on the $m_u - m_d$ mass difference itself) 
should not depend significantly on these details of the analysis, 
as long as both are carried out using the same method. 
We have determined $m^{\overline{\rm MS}}_+$ and
$m_s^{\overline{\rm MS}}$ in pure QCD using the leading order method
described here and find that the EM interactions tend to decrease these
masses by roughly one percent ($\sim O(\alpha_{\rm em})$) although it is
statistically less significant.
We are currently computing the EM splittings using PQ valence quarks and 
the conventional definition of $m_{res}$ to improve our determination of 
the quark masses.

\section{Discussion and prospects}
\label{sec:systematic errors}

The systematic errors in this calculation are quickly surveyed below.
In the measurements of neutral pions, we ignored the contribution from
the disconnected diagram,
which affects the determination of $\Delta_0^{(m)}$.
While it is possible to calculate this diagram explicitly, the clean
extraction of a signal is likely to be difficult.

In the study of the EM interactions, finite volume effects could be
significant as the photons are massless.
We have estimated the finite volume effects by considering the
vector-saturation model~\cite{Das:1967it,Bardeen:1988zw,Ecker:1988te}
and applying the physical volume of our lattices to this estimation. We find
roughly a $+10$\% increase in $\Delta_+^{(0)}$.
We expect a similar size of correction for the other new LECs, though
we need to check this.

Although it was not discussed in detail here, we also need to take into
account effects of the third dynamical quark and non-zero lattice spacing errors,
as well as effects of the quenched QED approximation.

 

\begin{thebibliography}{99} 

\bibitem{Izubuchi:lattice2005 review}
 For example, see
 the plenary review talks by T.~Izubuchi and M.~L\"uscher
 in these proceedings.

\bibitem{Duncan:1996xy}
 A.~Duncan, E.~Eichten and H.~Thacker, 
 \emph{Electromagnetic splittings and light quark masses in lattice QCD}, 
 Phys.\ Rev.\ Lett.\  {\bf 76}, (1996) 3894, 
 arXiv:hep-lat/9602005.

\bibitem{Kaplan:1992bt}
  D.~B.~Kaplan,
  \emph{A method for simulating chiral fermions on the lattice}, 
  Phys.\ Lett.\ B {\bf 288}, (1992) 342, 
  arXiv:hep-lat/9206013.

\bibitem{Shamir:1993zy}
  Y.~Shamir,
  \emph{Chiral fermions from lattice boundaries}, 
  Nucl.\ Phys.\ B {\bf 406}, (1993) 90, 
  arXiv:hep-lat/9303005.

\bibitem{Aoki:2004ht}
 Y.~Aoki {\it et al.} [RBC Collaboration],
 \emph{Lattice QCD with two dynamical flavors of domain wall fermions}, 
 arXiv:hep-lat/0411006. 

\bibitem{Hayakawa:2005eq}
 M.~Hayakawa, T,~Blum, T.~Izubuchi and N.~Yamada,
 \emph{Hadronic light-by-light scattering contribution to the muon g-2
 from lattice QCD: Methodology},
  PoS(LAT2005) 353, arXiv:hep-lat/0509016.

\bibitem{Suganuma}
 We thank H.~Suganuma for a useful comment on this point.

\bibitem{RBC NP Zm}
 The RBC Collaboration, in progress;
  C.~Dawson  [RBC Collaboration],
  \emph{Dynamical domain wall fermions},
  Nucl.\ Phys.\ Proc.\ Suppl.\  {\bf 128}, 54 (2004)
  [Nucl.\ Phys.\ Proc.\ Suppl.\  {\bf 129}, 167 (2004)]
  [arXiv:hep-lat/0310055].
 
\bibitem{Das:1967it}
  T.~Das, G.~S.~Guralnik, V.~S.~Mathur, F.~E.~Low and J.~E.~Young,
  \emph{Electromagnetic mass difference of pions}, 
  Phys.\ Rev.\ Lett.\  {\bf 18}, (1967) 759.

\bibitem{Bardeen:1988zw} 
  W.~A.~Bardeen, J.~Bijnens and J.~M.~Gerard,
  \emph{Hadronic matrix elements and the $\pi^+$-$\pi^0$ mass difference},
  Phys.\ Rev.\ Lett.\  {\bf 62}, (1989) 1343. 

\bibitem{Ecker:1988te} 
  G.~Ecker, J.~Gasser, A.~Pich and E.~de Rafael,
  \emph{The role of resonances in chiral perturbation theory}, 
  Nucl.\ Phys.\ B {\bf 321}, (1989) 311. 

\end{thebibliography}
\end{document}